# Which papers cited which tweets?
# An empirical analysis based on Scopus data

## Introduction

Much research has been published on the relationship between altmetrics (Priem, 2010; Priem, Taraborelli, Groth, & Neylon, 2010) and bibliometrics (Cronin & Sugimoto, 2014). Among all altmetrics, Twitter is one of the sources which have received most attention in scientometric research. Many studies reported which publications were mentioned in a tweet, and analyzed the correlation between citation counts of publications and the number of times they have been mentioned in a tweet (see for example: Liu, Lin, Xu, Shan, & Sheng, 2018; Uribe-Tirado & Alhuay-Quispe, 2017; Vogl, Scherndl, & Kuhberger, 2018). Bornmann (2015) performed a meta-analysis and found only a negligible correlation between twitter mentions and citation counts.

Subsequent studies also found weak or vanishingly small correlations between citation counts and twitter mentions (see for example Bornmann & Haunschild, 2018; de Winter, 2015; Snijder, 2016; Tonia, Van Oyen, Berger, Schindler, & Kunzli, 2016). The missing correlation of tweets and citations might show that tweets do not say anything about the usefulness, importance, impact etc. of publications. Tweets might be not more than noise without any signal for the interpretation of research activities. However, counting the number of tweets is already considered in the "Snowball Metrics Recipe Book" for measuring of social attention of research papers (Colledge, 2014).

In this study, we investigate the meaning of tweets for research evaluation from another perspective for discovering a possible policy relevance of tweets: we are interested in the tweets which were cited by publications and the publications which cited tweets. If many tweets were cited in publications, this might demonstrate that tweets have substantial and useful content.

## Methodology

We used the Scopus web-interface (http://www.scopus.com) to search for publications which cited tweets with the following query: 'REF("twitter.com/" PRE/2 "/status/") AND PUBYEAR < 2020'. Using this Scopus search query on 3 February 2020, we found 2910 publications published between 2007 and 2019. During the same period, more than 35 million papers were published. The meta-data of those publications were downloaded and analyzed using R (R Core Team, 2019).

The cited references from the 2910 papers were extracted and the occurrence of twitter.com followed by one or more characters of the following set was searched: letters between "a" and "z" (case-insensitive), numbers "0"-"9", "#", "!", ":", ";", "%", and "-". Finally, the string "status" was required. References matching such a pattern were found to contain tweet URLs (e.g., "https://www.twitter.com/marufins/status/1047467798996242433"). One paper has been found to not cite a single tweet (Scopus ID 2-s2.0-85055223597). This paper has been

Which papers cited which tweets? An empirical analysis based on Scopus data removed from the data set. Therefore, the remaining 2909 papers are analyzed in the following. The references containing tweet URLs were cleaned for further analysis. In total, 5506 cited tweets were found.

Results

Figure 1 shows the number of papers published per year which are citing tweets. Note that the papers from 2019 might not be fully indexed, yet. The average value per year is about 223.8. The numbers of papers citing tweets show an exponential increase. Recently, the number of papers citing tweets has doubled in about two years. 52.3% of the papers citing tweets so far were published in 2018 and 2019. For comparison, Haunschild, Bornmann, and Marx (2016) found that the climate change research literature doubled in five to six years and Bornmann and Mutz (2015) reported that the total volume of publications covered by the WoS between 1980 and 2012 doubled approximately every 24 years.

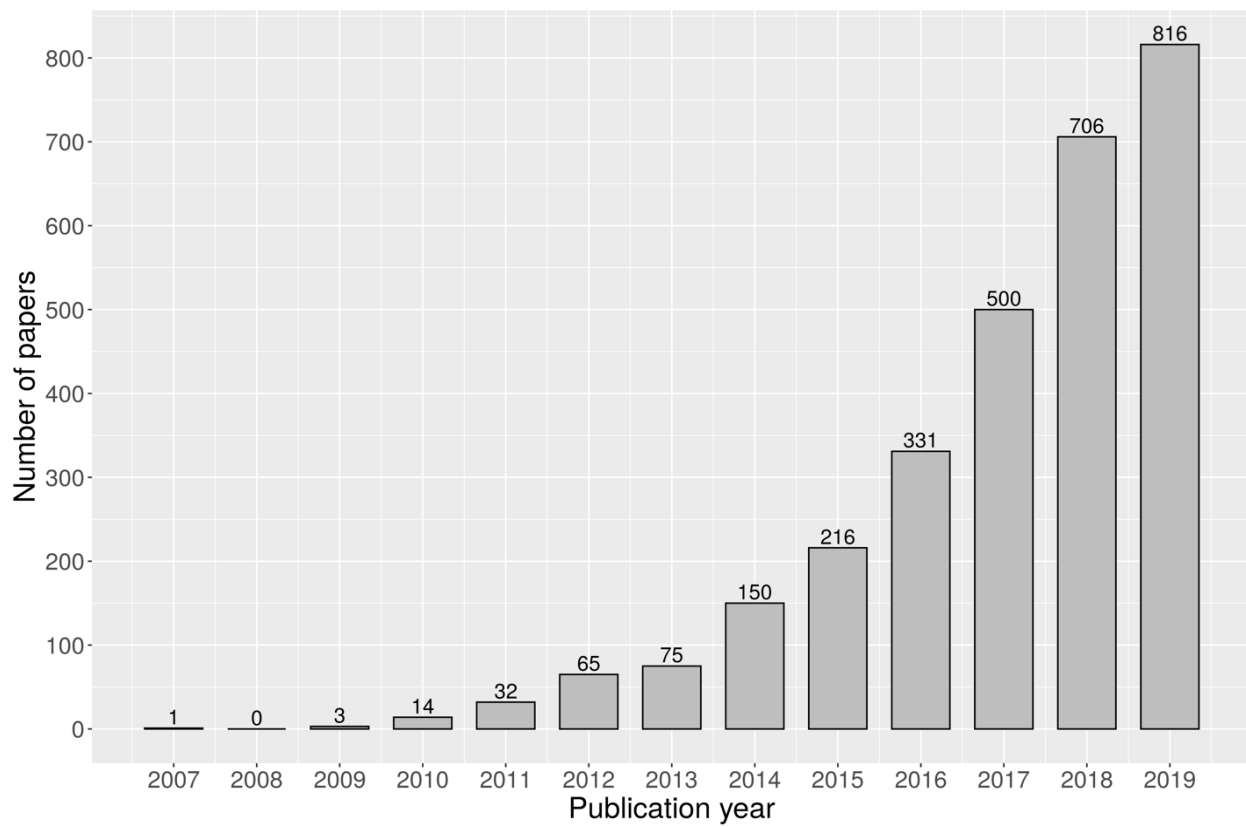

Figure 1: Number of papers published per year which are citing tweets

Figure 2 shows the distribution of the papers citing tweets across Scopus subject areas. Subject areas with less than 60 papers are aggregated in the slice "Others". The number of papers is given in parentheses. Note that many papers are classified in more than one subject area. In total, the paper set is broken down into 27 different subject areas; 10 of them are shown in Figure 2. It is rather surprising to see that most of the papers citing tweets are from Social Sciences, Arts and Humanities, and Computer Sciences. Those are disciplines which

Which papers cited which tweets? An empirical analysis based on Scopus data are not at the top of a ranking by the number of papers when all Scopus-indexed papers are analyzed.

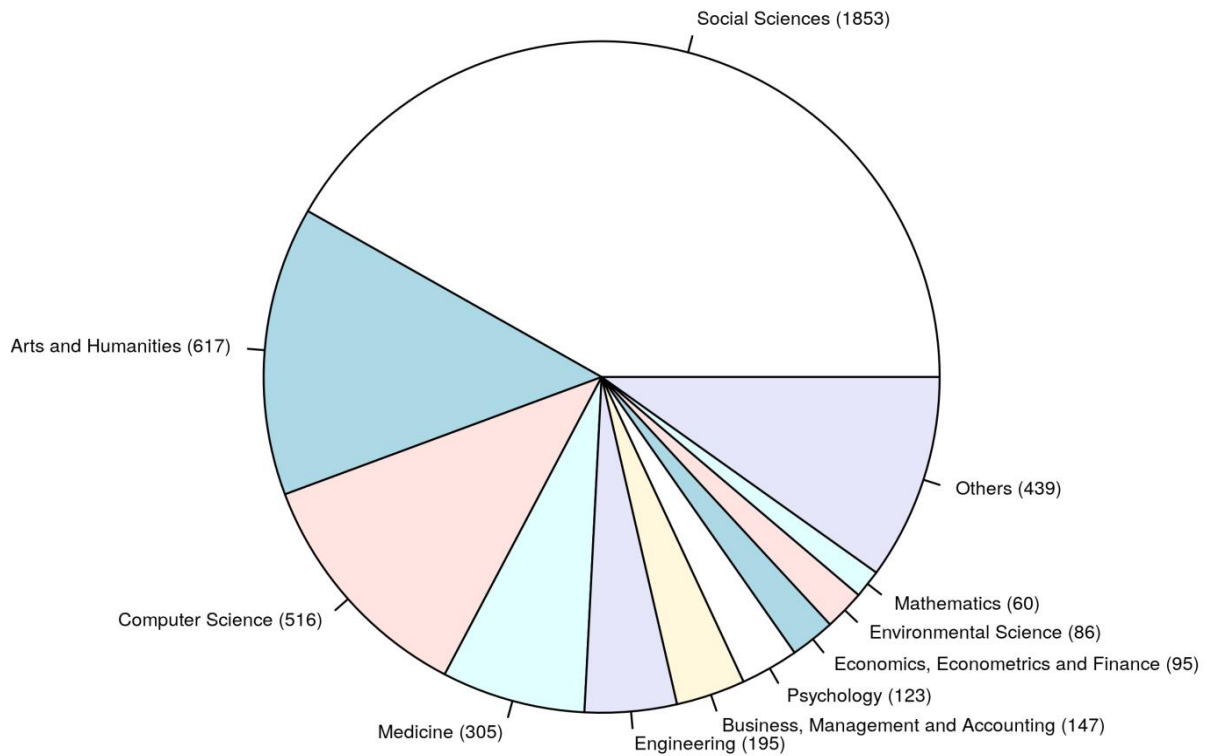

Figure 2: Analysis of Scopus subject areas of papers which cited tweets

74.25% of the papers (n= 2160) cited only a single tweet. Figure 3 shows the frequency distribution of cited tweets in papers for papers which have cited at least two tweets. The paper which cited most tweets (DOI 10.1080/15456870.2019.1610763) has analyzed the Twitter usage of Donald Trump: it examines "how Donald Trump has used Twitter to separate himself rhetorically from the traditional Washington elite and maintain his outsider status while serving as president" (p. 183). Although this paper has cited 55 tweets, the tweets have not been cited because they had cognitive influence, but they were study object. It is clear that this kind of impact produced by tweets is not useful for research evaluation.

Which papers cited which tweets? An empirical analysis based on Scopus data

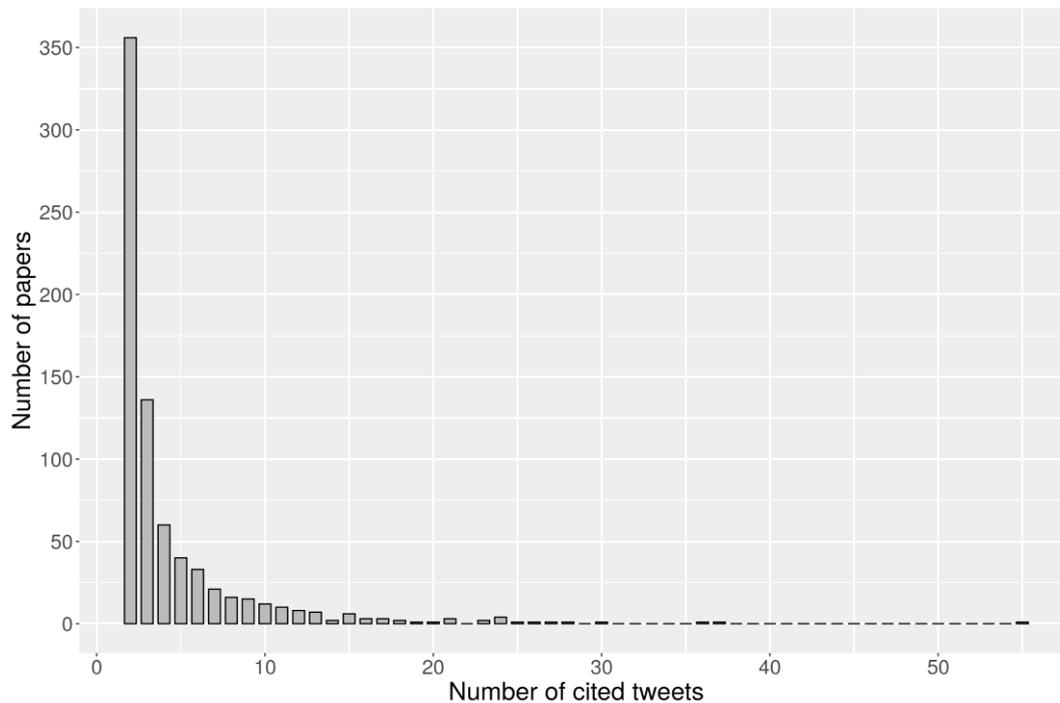

Figure 3: Frequency distribution of cited tweets in papers for papers which have cited at least two tweets

Table 1 shows the tweets which were cited at least five times. The most cited tweet has been cited 17 times so far. It is the rather famous tweet by Donald Trump ("The concept of global warming was created by and for the Chinese in order to make U.S. manufacturing non-competitive.") from 06 November 2012. The second most cited tweet with 14 citations is from Jason Priem with which he coined the term altmetrics on 29 September 2010: "I like the term #articlelevelmetrics, but it fails to imply *diversity* of measures. Lately, I'm liking #altmetrics". The citations of this tweet might reflect cognitive influence, since Jason Priem proposed the name of a new type of indicators (that are very popular today) (Tahamtan & Bornmann, 2018). The third most cited tweet with eleven citations is by Alyssa Milano in the context of the "me too" movement. The tweet number 7 in Table 1 was the very first tweet ever posted.

Table 1: Most frequently cited tweets

| Nº | Twitter URL (omitting the twitter base URL) | Number of citations |
|---|---|---|
| 1 | realdonaldtrump/status/265895292191248385 | 17 |
| 2 | jasonpriem/status/25844968813 | 14 |
| 3 | alyssamilano/status/919659438700670976 | 11 |
| 4 | realdonaldtrump/status/832708293516632065 | 10 |
| 5 | realdonaldtrump/status/890196164313833472 | 8 |
| 6 | realdonaldtrump/status/890193981585444864 | 7 |
| 7 | jack/status/20 | 7 |
| 8 | twitter/status/281051652235087872 | 7 |
| 9 | chrismessina/status/223115412 | 6 |

Which papers cited which tweets? An empirical analysis based on Scopus data

| 10 | olesovhcom/status/778830571677978624 | 6 |
| 11 | internetpartynz/lists/status/451000000000000000 | 5 |
| 12 | kohenari/status/1057989081156730881 | 5 |
| 13 | realdonaldtrump/status/827867311054974976 | 5 |
| 14 | realdonaldtrump/status/872086906804240384 | 5 |
| 15 | jkrums/status/1121915133 | 5 |
| 16 | nsf/statuses/129637763745189889 | 5 |
| 17 | realdonaldtrump/status/828342202174668800 | 5 |
| 18 | realdonaldtrump/status/890197095151546369 | 5 |

Figure 4 shows the most frequently cited twitter handles ("@" signs are omitted) with the number of citations in parentheses. Only the twitter handles cited at least five times are shown. The size of the letters also reflects the citation counts. The most cited twitter handles are "realdonaldtrump" with 514 citations followed by "hillaryclinton" with 48 citations and "narendramodi" with 35 citations. This indicates a high proportion of political science studies in the paper set.

Figure 4: Most frequently cited twitter handles with the number of citations in parentheses

Which papers cited which tweets? An empirical analysis based on Scopus data

Discussion

Twitter is a very popular altmetric for measuring broader impact. For example, the company Altmetric monitors this source for its so-called altmetric attention score (reflecting online activity and discussion). In this study, we are interested in the question whether the use of tweets is justified in research evaluation: are tweets of interest for authors of research papers?

We found only 2909 papers which have cited at least one tweet. Papers which cited tweets are currently increasing at an exponential rate. The number of publications has doubled in the most recent two years. Most of the papers citing tweets are from the subject areas Social Sciences, Arts and Humanities, and Computer Sciences. Most of the papers cited only a single tweet. Up to 55 tweets cited in a single paper were found. The three most cited tweets are Donald Trump's false claim that climate change was an invention by China, Jason Priem's proposal of the term altmetrics, and a comment by Alyssa Milano in the context of the "me too" movement.

Scientific impact is measured by citation counts. According to Seglen (1992), around 45% of the papers published annually and indexed in the Science Citation Index receive at least one citation within five years after publication. Using the same proxy, we can conclude that rather few tweets have gained scientific impact, especially considering that around 6000 tweets are posted each second (see https://www.internetlivestats.com/twitter-statistics/). Our results, therefore, confirm the results by Bornmann (2015) and others that tweets are rather noise than signal. Tweet counts do not seem to be useful for research evaluation purposes.